\documentclass[iop]{emulateapj}
\usepackage{color,ulem}
\usepackage{epsf}
\usepackage{epsfig}
\usepackage{amsmath}
\usepackage{amssymb}
\usepackage{graphicx}
\usepackage{booktabs}

\usepackage{pifont}
\usepackage{txfonts}

\shorttitle{NDAF with Pm-dependent $\alpha$}
\shortauthors{Kawanaka \& Masada}

\begin{document}


\title{Neutrino-Dominated Accretion Flows with Magnetic Prandtl Number-Dependent MRI-driven Turbulence}


\author{Norita Kawanaka}
\affil{Department of Astronomy, Graduate School of Science, Kyoto University, Kitashirakawa Oiwake-cho, Sakyo-ku, Kyoto 606-8502, Japan}
\affil{Hakubi Center, Kyoto University, Yoshida Honmachi, Sakyo-ku, Kyoto, 606-8501, Japan}
\email{norita@kusastro.kyoto-u.ac.jp}

\author{Youhei Masada}
\affil{Department of Physics and Astronomy, Aichi University of Education; 1 Hirosawa, Igaya-cho, Kariya, Aichi 448-8542, Japan}

\begin{abstract}
We investigate the stability of a neutrino-dominated accretion flow (NDAF), which is expected to be formed in the gravitational collapse of a massive star or the merger of a neutron star binary, based on the variable-$\alpha$ prescription.  Recent magnetohydrodynamic (MHD) simulations shows that the viscosity parameter $\alpha$ is proportional to the power of the magnetic Prandtl number ${\rm Pm}=\nu/\eta$, where $\nu$ and $\eta$ are the kinematic viscosity and electric resistivity of the fluid, respectively.  In the inner region of a hyperaccretion flow, the viscosity and resistivity are carried by mildly, relativistically degenerated electrons.  We fit the dependence of the magnetic Prandtl number on density and temperature by a simple analytic form, and derive the condition for an NDAF to be dynamically unstable.  As demonstrations we perform simple one-dimensional simulations of NDAFs with the variable-$\alpha$ and show that the mass accretion becomes highly time-variable in the unstable branch.  This mechanism may account for the rapid variability observed in the prompt emission of gamma-ray bursts (GRBs).  The mass ejection from a hyperaccretion flow due to viscous heating, which makes a kilonova/macronova emission in the merger of a neutron star binary, is also briefly discussed.
\end{abstract}

\section{Introduction}
It is widely believed that gamma-ray bursts (GRBs) are powered by hyperaccreting black holes (or neutron stars, possibly) which are formed after the gravitational collapse of a massive star or the compact binary (neutron star - neutron star, or neutron star - black hole) merger \citep{1992ApJ...395L..83N, popham+99, 2001ApJ...557..949N}.  The accretion rate would be as high as $0.01-1M_{\odot}~{\rm s}^{-1}$, and due to its high density ($\sim 10^{9-10}~{\rm g}~{\rm cm}^{-3}$) and temperature ($\sim 10^{9-11}~{\rm K}$), the accretion flow is mainly cooled via neutrino emission.  It is often called a neutrino-dominated accretion flow (NDAF), and its structure and behavior have been investigated by many researchers (\citealp{2002ApJ...577..311K,2002ApJ...579..706D,2005ApJ...629..341K,2006ApJ...643L..87G,2007ApJ...657..383C,kawanakamineshige07,2007ApJ...661.1025L,2009MNRAS.397.1153K,2013ApJ...765..125L,2013ApJ...766...31K,2013ApJS..207...23X,2015ApJ...805...37L,2017ApJ...849...47L, 2017ApJ...837...39J}; for the recent comprehensive review see \citealp{2017NewAR..79....1L}).  In this model, it is considered that an ultrarelativistic jet is launched from this accretion system due to the energy deposition process such as neutrino pair annihilation \citep{1989Natur.340..126E, 2000ApJ...531..949A,2011MNRAS.410.2302Z,2013MNRAS.428.2443S} or magnetohydrodynamical mechanisms such as Blandford-Znajek process \citep{1977MNRAS.179..433B, 2004ApJ...611..977M, 2006ApJ...641..103H, 2013ApJ...766...31K}.

Observations show that the prompt emission of GRBs is highly variable, and its typical timescale is as short as $10^{-3}-10^{-1}~{\rm s}$ (e.g. \citealp{2013MNRAS.432..857M}).  Although the origin of such violent behavior is still uncertain, it is often considered to be driven by a certain instability of the NDAF \citep{kawanakamineshige07,2007ApJ...664.1011J,masada+07,2012MNRAS.419..713K,2013ApJ...777L..15K,2015PASJ...67..101K}.  When a disk becomes unstable, violent and sporadic mass accretion is expected, which would launch a highly variable jet from the vicinity of a central black hole.  Most of the works about the disk instability in NDAFs (except for \citealp{masada+07}) are based on the $\alpha$-viscosity model proposed by \cite{1973A&A....24..337S}, where the ratio of the viscous stress to the isotropic pressure in an accretion disk is constant.  This model has been widely accepted and used to explain the liberation of gravitational energy and behavior such as outbursts occurring in accretion disk systems such as dwarf novae, X-ray binaries, and active galactic nuclei (see \citealp{2008bhad.book.....K} for the detail).  \cite{1991ApJ...376..214B} pointed out the possibility that the viscosity in accretion disks comes from the magnetohydrodynamic (MHD) turbulence operated by the magnetorotational instability (MRI).  Since their indication, the non-linear growth of MRI and the turbulent angular momentum transport have been investigated in many studies \citep{1992ApJ...400..595H, 1995ApJ...440..742H, 2004ApJ...605..321S, 2007A&A...476.1123F}.  In this context, the shear viscosity in an accretion disk is determined by the saturation level of the growth of MRI, which is assumed to be simply determined by the value of isotropic pressure in the $\alpha$-viscosity model.

According to the recent numerical studies, the saturated value of the viscosity depends on the magnetic Prandtl number, ${\rm Pm}=\nu/\eta$, where $\nu$ and $\eta$ are the kinetic viscosity and magnetic diffusivity, respectively \citep{2007A&A...476.1123F,2007MNRAS.378.1471L,2009ApJ...707..833S,2010A&A...516A..51L,2011ApJ...730...94S}.  Especially, when the net magnetic flux is non-zero, the effective value of $\alpha$-viscosity is found to be proportional to the power of the magnetic Prandtl number, $\alpha\propto {\rm Pm}^{\beta}$, where $\beta$ is a positive number \citep{2007A&A...476.1123F,2009ApJ...707..833S}.  Taking into account the effect of ${\rm Pm}$-dependence of $\alpha$-viscosity, the stability criteria for an accretion disk would be modified.  For example, \cite{takahashi11} constructed the steady state model for a geometrically thin, optically thick disk using ${\rm Pm}$-dependent $\alpha$-viscosity and found that the disk becomes gravitationally, thermally, or secularly unstable for a sufficiently large power-law index $\beta$.  More recently, \cite{potter14} investigated the spectrum and time-dependent behavior of such an unstable disk and found that they are analogous to what are observed in flaring X-ray binary systems (see also \citealp{potter17}).

In this paper we firstly discuss the stability of an NDAF with Pm-dependent $\alpha$-viscosity, and investigate its time-dependent behavior by one-dimensional simulations of an NDAF.  In evaluating the magnetic Prandtl number in NDAFs, we use the results presented in \cite{2008MNRAS.391..922R}, which discussed the viscosity and magnetic diffusivity in an NDAF taking into account the degeneracy of electrons.  In Section 2 we construct an analytic model of an NDAF using variable $\alpha$-viscosity prescription and evaluate the condition that an NDAF becomes unstable.  In Section 3 we present the time evolution of an NDAF with variable $\alpha$-viscosity and show that the mass accretion is highly time-variable in the unstable branch.  Section 4 is devoted to the interpretation of our results and application to the prompt emission of GRBs, which show highly violent behavior.  We summarize this work in Section 5.

\section{Analysis}
\subsection{Fundamental Equations of an NDAF}
In this section we construct the model of a neutrino-cooled accretion disk taking into account the Pm-dependence of $\alpha$, and investigate its stability.  The vertically-integrated equations governing the structure of a neutrino-cooled accretion disk are the following (see \citealp{kawanakamineshige07,2013ApJ...766...31K} and references therein):

1. Mass conservation equation:
\begin{eqnarray}
\dot{M}=-2\pi r \varv_r \Sigma, \label{masscons}
\end{eqnarray}

2. Kepler rotation (Newtonian gravity):
\begin{eqnarray}
\Omega=\sqrt{\frac{GM_{\rm BH}}{r^3}}, \label{kepler}
\end{eqnarray}

3. Hydrostatic balance:
\begin{eqnarray}
\frac{p}{\rho}=\Omega^2 H^2, \label{hydrostatic}
\end{eqnarray}

4. Angular momentum conservation:
\begin{eqnarray}
-\nu_t\Sigma \frac{d\Omega}{d{\ln}r}=\frac{\dot{M}\Omega}{2\pi}\left( 1-\sqrt{\frac{R_{\rm in}}{r}} \right), \label{angcons}
\end{eqnarray}

5. Energy balance:
\begin{eqnarray}
Q^+\equiv \nu_t \Sigma \left( \frac{d\Omega}{d\ln r} \right)^2 = Q^-, \label{energybalance}
\end{eqnarray}

6. $\alpha$ prescription:
\begin{eqnarray}
-\nu_t\Sigma \frac{d\Omega}{d\ln r} = \alpha p H. \label{alpha}
\end{eqnarray}
Here $\varv_r$, $\Sigma$, $\nu_t$, $R_{\rm in}$, $Q^-$, $p$, $\rho$ and $H$ are the radial velocity, the surface density, the kinematic viscosity due to the turbulence driven by MRI, the inner radius of the disk, the cooling rate per unit surface area, the disk pressure, the mass density of the disk, and the disk scale height, respectively.  The surface density is defined as $\Sigma=2\rho H$. 

The cooling rate $Q^-$ is in general described by the sum of two components:
\begin{eqnarray}
Q^-=Q_{\rm adv}^-+Q_{\nu}^-,
\end{eqnarray}
where $Q_{\rm adv}^-=T\Sigma \varv_r ds/dr$ is the advective cooling rate, where $s$ is the specific entropy and $T$ is the temperature in the disk midplane, and $Q_{\nu}^-$ is the cooling due to neutrino emission from the disk.  The most effective emission process for electron-type neutrinos in an NDAF is the URCA process ($e^-/e^+$-capture onto a proton/neutron), whose emissivity (i.e. energy loss rate per unit volume) can be approximated as
\begin{eqnarray}
q_{eN}&=&9.0\times 10^{33}~{\rm erg}~{\rm cm}^{-3}~{\rm s}^{-1}\rho_{10}T_{11}^6,
\end{eqnarray}
where $\rho_{10}=\rho/10^{10}~{\rm g}~{\rm cm}^{-3}$ and $T_{11}=T/10^{11}~{\rm K}$ \citep{2002ApJ...579..706D}.  As for the other types of neutrinos (i.e., mu- and tau-type neutrinos), electron-positron pair annihilation ($e^- + e^+ \rightarrow \nu_x+\bar{\nu}_x$) is dominant, and its emissivity is approximately described as
\begin{eqnarray}
q_{e^-e^+\rightarrow \nu_x\bar{\nu}_x}=1.4\times 10^{33}~{\rm erg}~{\rm cm}^{-3}~{\rm s}^{-1}T_{11}^9,
\end{eqnarray}
which is valid in the nondegenerate limit, and when electrons are strongly degenerated we can neglect these processes.

The neutrino cooling rate can be described by the two-stream approximation \citep{2002ApJ...579..706D,2005ApJ...629..341K}:
\begin{eqnarray}
Q_{\nu}^-=2\sum_i \frac{(7/8)\sigma T^4}{(3/4)(\tau_{\nu_i}/2+1/\sqrt{3}+1/(3\tau_{a,\nu_i}))}, \label{nucoolingrate}
\end{eqnarray}
where $\tau_{a,\nu_i}$ and $\tau_{s,\nu_i}$ represent absorptive and scattering optical depths, respectively, and $\tau_{\nu_i}=\tau_{a,\nu_i}+\tau_{s,\nu_i}$ is the sum of absorptive and scattering optical depth with respect to $\nu_i$ ($=\nu_e,\bar{\nu}_e$).  From the Kirchhoff's law, the absorptive optical depth is given by
\begin{eqnarray}
\tau_{a,\nu_i}&=&\frac{q_{\nu_i}^-H}{4(7/8)\sigma T^4} \nonumber \\
&\simeq& 4.5\times 10^{-7}T_{11}^2\rho_{10}H,
\end{eqnarray}
and the scattering optical depth, which is dominated by nucleons, is given by
\begin{eqnarray}
\tau_{s,\nu_i}\simeq 2.7\times 10^{-7}T_{11}^2\rho_{10}H,
\end{eqnarray}
both of which have the same temperature and density dependence \citep{2002ApJ...579..706D}.

The pressure in a disk, $p$, is described as the sum of the contributions from radiation, baryonic gas, degenerate electrons, and neutrinos (if they are trapped):
\begin{eqnarray}
p\simeq \frac{11}{12}aT^4+\frac{\rho k_B T}{m_p}+\frac{2\pi hc}{3}\left( \frac{3}{8\pi m_p} \right)^{4/3}\left( Y_e\rho \right)^{4/3}+\frac{u_{\nu}}{3},
\end{eqnarray}
where $Y_e=n_p/(n_p+n_n)$ is the electron fraction, $a\simeq 7.56\times 10^{-15}~{\rm erg}~{\rm cm}^{-3}~{\rm deg}^{-4}$ is the radiation constant, and $k_B$ is the Boltzmann constant.

When the mass accretion rate is the order of $\sim 0.01M_{\odot}~{\rm s}^{-1}$, the cooling rate $Q^-$ is dominated by the neutrino emission via URCA process, the optical depth of the disk with respect to neutrinos is thin, and the disk pressure is dominated by baryonic gas in the inner disk region (see \citealp{kawanakamineshige07} and references therein).  Then we can approximate the disk pressure and the neutrino energy flux as
\begin{eqnarray}
p&\approx &\frac{\rho k_B T}{m_p}, \label{eos} \\
Q^-&\approx &9.0\times 10^{33}~{\rm erg}~{\rm cm}^{-2}~{\rm s}^{-1}\rho_{10}T_{11}^6 H, \label{urcathin}
\end{eqnarray}
and when the mass accretion rate is the order of $\sim 0.1M_{\odot}~{\rm s}^{-1}$, the disk pressure is still dominated by baryonic gas but the disk becomes optically-thick with respect to neutrinos and we can approximate the neutrino energy flux as
\begin{eqnarray}
Q^-\approx 7.4\times 10^{46}~{\rm erg}~{\rm cm}^{-2}~{\rm s}^{-1}T_{11}^2\rho_{10}^{-1}H^{-1}, \label{urcathick}
\end{eqnarray}
where $\rho_{11}\equiv \rho/10^{11}~{\rm g}~{\rm cm}^{-3}$ and $T_{10}\equiv T/10^{10}~{\rm K}$ \citep{popham+99,2013ApJ...766...31K}.  

\subsection{Instability Criteria}
Let us derive the stability criterion for this hyperaccretion flow.  The condition for the secular instability can be described as
\begin{eqnarray}
\frac{\partial \left({T_{r\phi} H}\right)}{\partial \Sigma}<0, \label{unstable}
\end{eqnarray}
where $T_{r\phi}=\rho \nu_t (d\Omega/d\ln r)$ is the shear viscous stress.  The physical meaning of this condition is the following.  When this inequality is fulfilled, a local increase of the surface density gives rise to the local decrease of the shear stress, which would suppress the local mass accretion rate.  Then mass would accumulate locally and the surface density grows further.  In this sense the initial perturbation given in an accretion flow would be amplified when Eq. (\ref{unstable}) is satisfied.  

Let us evaluate the condition for the secular instability in a disk with optically-thin neutrino cooling.  From Eqs. (\ref{energybalance}), (\ref{alpha}), (\ref{eos}) and (\ref{urcathin}) we obtain
\begin{eqnarray}
\alpha \propto T^5, \label{prop1}
\end{eqnarray}
and from Eqs. (\ref{hydrostatic}) and (\ref{eos}) we obtain
\begin{eqnarray}
T\propto H^2,\label{prop2}
\end{eqnarray}
at a fixed radius.  Then we can rewrite the instability condition (\ref{unstable}) as
\begin{eqnarray}
\frac{\partial \left( \alpha^{6/5}\Sigma\right)}{\partial \Sigma}<0, \label{unstable2}
\end{eqnarray}
at the fixed radius.  On the other hand, the instability criterion for a disk with optically-thick neutrino cooling is obtained from Eqs. (\ref{energybalance}), (\ref{alpha}), and (\ref{urcathick}) as
\begin{eqnarray}
\frac{\partial \left(\alpha^2 \Sigma^3 \right)}{\partial \Sigma}<0,
\end{eqnarray}
at a fixed radius.  Note that if $\alpha$ is constant and independent of density or temperature, these conditions are never fulfilled, which means that the accretion flow would be stable as long as its cooling is dominated by neutrino emission due to the URCA process.

Here we evaluate the magnetic Prandtl number, ${\rm Pm}=\nu/\eta$, in an NDAF to find the disk instability criterion.  In \cite{2008ApJ...674..408B} they consider the microscopic kinematic viscosity $\nu$ is dominated by the Coulomb scattering or radiation, and the resistivity $\eta$ by the Coulomb scattering \citep{spitzer62,1984frh..book.....M}.  However, these values are not applicable to NDAFs because they assume that electrons are non-relativistic and non-degenerated, which is generally not the case in NDAFs.  \cite{2008MNRAS.391..922R} investigated the microscopic physical properties in NDAFs taking into account the relativistic degeneracy of electrons and found that the viscosity and resistivity are modified from the values adopted \cite{2008ApJ...674..408B}.  We fit the viscosity and resistivity presented in Fig. 3 of \cite{2008MNRAS.391..922R} by simple analytic functions and obtain the fitting formula for the magnetic Prandtl number as
\begin{eqnarray}
{\rm Pm}\simeq 4.5\times 10^2 T_{11}^2 \rho_{10}^{-1}. \label{pmfit}
\end{eqnarray}
Then we can rewrite our assumption on the dependence of the viscosity parameter on the magnetic Prandtl number, $\alpha \propto {\rm Pm}^{\beta}$, as
\begin{eqnarray}
\alpha \propto \rho^{-\beta}T^{2\beta}. \label{prandtlrhotemp}
\end{eqnarray}
Using Eqs. (\ref{prop1}), (\ref{prop2}) and (\ref{unstable2}), we can evaluate the criterion for $\beta$ that the accretion disk cooled dominantly by neutrino emissions would suffer from the secular instability.  In the case when the disk is optically-thin, the criterion is
\begin{eqnarray}
\frac{10}{17}<\beta<2, \label{thinunstable}
\end{eqnarray}
while in the case when the disk is optically-thick, the criterion reads as
\begin{eqnarray}
\beta > \frac{2}{5}. \label{thickunstable}
\end{eqnarray}

\subsection{Disk Evolution Scenario}
As shown in the previous section, a hyperaccretion flow can be secularly unstable when the index $\beta$ is within a certain range.  When the instability sets in locally, the mass would be accumulated in the unstable zone and the surface density of a disk grows locally.  As a result, the accretion flow would be gravitationally unstable and the gravitational torque would become effective.  This causes the intense mass accretion onto a black hole from that zone.  Just after that the local surface density becomes lower but as the mass accretion from the outer part continues that zone would be filled again and become secularly unstable again.  In this way the intense mass accretion onto a black hole occurs repeatedly, which may be the origin of the short-term variability observed in the prompt emission of GRBs.

The suppression of MRI-driven turbulence and following evolution of an accretion flow have been discussed in the context of protoplanetary disks \citep{nakano91, jin96, gammie96, sanomiyama99, sano+00, armitage+01}, in which the Ohmic dissipation of the electric current in a disk can suppress the growth of MRI.  In the context of a hyperaccretion flow as a central engine of GRBs, \cite{masada+07} discuss the suppression of MRI-driven turbulence due to neutrino diffusion in a neutrino-thick hyperaccretion flow whose mass accretion rate is as high as $\sim 1M_{\odot}~{\rm s}^{-1}$, followed by intermittent mass accretion driven by gravitational instability.  However, in the current model it is not necessary for a hyperaccretion flow to be optically-thick with respect to neutrinos, as shown above.  As long as an accretion flow is dominantly cooled via neutrino emissions the secular instability and subsequent intermittent accretion can appear for a certain range of $\beta$.  In the next section, we present the results of one-dimensional simulations of a hyperaccretion flow according to this evolution scenario.

\section{1D Simulation of Disk Evolution} 

\subsection{Basic Equation and Simulation Setup}
We perform one-dimensional numerical simulations as a demonstration of the dynamical evolution of the NDAF with Pm-dependent
$\alpha$-viscosity.  The basic equations are the mass and angular momentum conservation in the disk, 
\begin{eqnarray}
  \frac{\partial \Sigma}{\partial t}=\frac{3}{r}\frac{\partial}{\partial r}\left[ r^{1/2}\frac{\partial}{\partial r}
    \left( \nu_t \Sigma r^{1/2} \right) \right]\;, \label{evolution}
\end{eqnarray}
and %the equation of 
the energy conservation,
\begin{eqnarray}
\frac{\partial \ln T}{\partial t}+v_r\frac{\partial \ln T}{\partial r} 
&=&-\frac{4-3\chi}{(12-21\chi/2)}\frac{1}{r}\frac{\partial}{\partial r}\left( rv_r\right) \nonumber \\
&&+\frac{1}{2P H}\frac{1}{(12-21\chi/2)}\left(Q^+-Q^- \right) \;, \label{energy} 
\end{eqnarray}
where $\chi$ is the ratio of gas pressure to total pressure, $Q^+ \equiv \nu_t \Sigma ({\rm d}\Omega/{\rm d}\ln r)^2$
is the viscous heating rate, and $Q^-$ is the neutrino cooling rate described by eq.(\ref{nucoolingrate}).  The radial velocity of the accretion flow is given by
\begin{eqnarray}
v_r=-\frac{3}{\Sigma r^{1/2}}\frac{\partial}{\partial r}\left[ \nu_t \Sigma r^{1/2} \right]. \label{velocity}
\end{eqnarray}
The advection terms in equation~(\ref{energy}) are neglected in our actual simulation for the simplicity
since it can be expected that they have little impact on the energy balance in the relatively small mass accretion regime
($\dot{M} = \mathcal{O}(10^{-2})\ M_\odot/{\rm sec}$) of our interest. 

We take into account the Pm-dependence of $\alpha$ in the similar way as Potter \& Balbus (2014):
\begin{eqnarray}
\alpha ({\rm Pm}) =\alpha_{\rm min}+(\alpha_{\rm max}-\alpha_{\rm min})\left(\frac{{\rm Pm}^{\beta}}{{\rm Pm}^{\beta}+{\rm Pm}_c}\right). \label{alphaparam}
\end{eqnarray}
While the ceiling value of $\alpha$ is fixed to $\alpha_{\rm max}=0.1$, the floor value $\alpha_{\rm min}$ is remained as a control parameter
of our simulation.  This functional form bridges the small and large Pm regimes smoothly, and behaves approximately as being proportional
to ${\rm Pm}^\beta$ when ${\rm Pm} \lesssim {\rm Pm}_c$. Note that ${\rm Pm}$ is given, as a simple function of $\rho$ and $T$,
by eq.(22) in this study.  The critical value ${\rm Pm}_c$ is varied depending on the simulation model. 

As stated in \S~2.2, when a part of an accretion disk becomes secularly unstable, mass would accumulate in that part and
the surface density locally grows with time, resulting in the formation of the gravitationally unstable region at some
evolutionary stage.

In our simulations, the condition for the onset of the gravitational instability is given, with Toomre's $\mathcal{Q}$ parameter, by
\begin{eqnarray}
\mathcal{Q}\equiv\frac{c_s \Omega}{\pi G \Sigma} < \mathcal{Q}_{\rm crit} \;, \label{toomre}
\end{eqnarray}
where $c_s=(p/\rho)^{1/2}$ is the speed of sound and $\mathcal{Q}_{\rm crit}$ is the critical value for the instability which is evaluated to be in the
range $1 \lesssim \mathcal{Q}_{\rm crit} \lesssim 5$ from the numerical simulation of the accretion disk.  Once an accretion flow becomes
gravitationally unstable, non-axisymmetric structure and hence the gravitational torque arises, leading to the outward angular momentum transport
in the disk.  To model this process, we simultaneously solve the evolution equation of $\alpha_{\rm grav}$
($\alpha$-parameter due to the gravitational instability), similar to the one proposed by Zhu et al. (2010),
\begin{eqnarray}
  \frac{\partial \alpha_{\rm grav}}{\partial t}=-\sigma\frac{\alpha_{\rm grav}^2-\alpha_0^2}
       {\alpha_{\rm grav}^2+\alpha_0^2}+\nu_{g}\frac{\partial^2 \alpha_{\rm grav}}{\partial r^2} \;, \label{alphagrav}
\end{eqnarray}
where $\alpha_0 = 0.1$ is the reference value, $\sigma$ is a control parameter for the evolution of $\alpha_{\rm grav}$ and chosen
so as to match the typical growth rate of the gravitational instability, i.e., $\sigma=\Omega\sqrt{(\mathcal{Q}_{\rm crit}/\mathcal{Q})^2-1}$. The last
term of the RHS represents the turbulent diffusion of $\alpha_{\rm grav}$ and the coefficient $\nu_{g}$ is assumed to be the product
of the typical size and velocity of the turbulence, i.e., $\nu_{g}= c_s^2/\Omega$ based on the mixing-length theory.

Computationally, we solve equations (\ref{evolution}), (\ref{energy}), (\ref{velocity}), and (\ref{alphagrav}) with explicit finite difference method, and then include the self-gravity as an additional
sweep over the grid using $\nu_{\rm grav} \equiv \alpha_{\rm grav}c_s^2/\Omega$ and the surface density $\Sigma$, and as an additional heating
term in the energy equation when the condition (\ref{toomre}) is satisfied. The computational grid of $256$ radial mesh-points is uniform in a scaled
radial variable $x \propto r^{1/2}$.

For the numerical setup, we adopt the mass of the central black hole $M_{\rm BH} = 3M_\odot$, an inner disk radius $r_{\rm in} = 3r_g$,
and an outer disk radius $r_{\rm out} = 50r_g$, where $r_g = GM_{\rm BH}/c^2 = 4.5\times 10^5$ cm is the Schwarzschild radius.  A zero-torque boundary condition is imposed at $r_{\rm in}$, while at $r_{\rm out}$ we impose two types of 
boundary conditions: (1) the constant mass accretion rate, i.e., $\dot{M}_{\rm out} = {\rm const.}$ (\S~3.2), and (2) 
the time-decreasing mass accretion rate, i.e., $\dot{M}_{\rm out} \propto t^{-5/3}$ (\S~3.3), which is expected in the fallback accretion just
after the failed supernova explosion or the merger of a neutron star binary. For the initial profiles of $\Sigma$ and $T$, the analytic
solutions of equations (\ref{masscons})--(\ref{alpha}), and (\ref{nucoolingrate}) with constant mass accretion rate $\dot{M}$ and constant
viscous parameter $\alpha$ are adopted \citep{2013ApJ...777L..15K}.  After the relaxation to
the equilibrium state with the constant $\alpha$, we take into account the Pm-dependence of $\alpha$ and restart the time-dependent calculation.

\subsection{Models with Constant Mass Injection}
%%%%%%%%%%%%%%%%%%%%%%%%%%%%%%%%%%%%%%%%%%%%%%%%%%%%%%%%%%%%%%%%%%%%
\begin{figure*}[tbp]
\begin{center}
\includegraphics[width=17cm,clip]{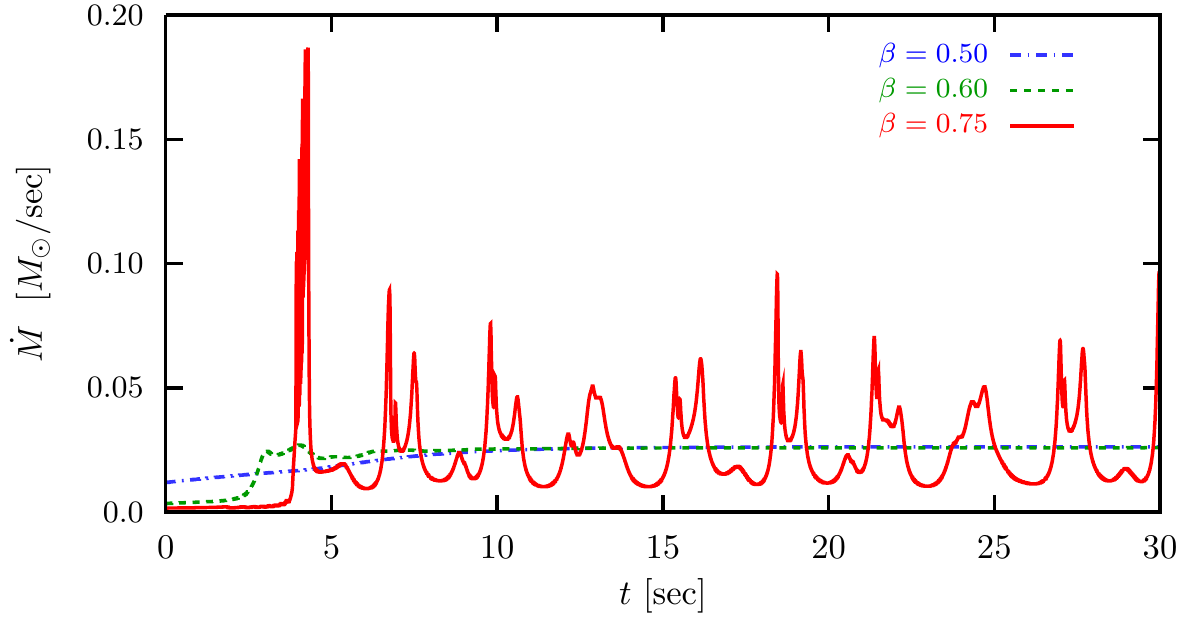}
\caption{Temporal evolution of $\dot{M}$ in unit of $M_\odot {\rm /sec}$ for the
models with $\beta = 0.5$ (blue), $\beta = 0.6$ (green) and $\beta = 0.75$ (red).  The mass injection rate at the outer boundary is $\dot{M}_{\rm out}\simeq 0.025M_{\odot}/{\rm sec}^{-1}$.  Other parameters are adopted as $\alpha_{\rm min}=10^{-4}$, ${\rm Pm}_{\rm c}=10$, and $\mathcal{Q}_{\rm crit} = 5.0$.}
\label{fig1}
\end{center}
\end{figure*}
%%%%%%%%%%%%%%%%%%%%%%%%%%%%%%%%%%%%%%%%%%%%%%%%%%%%%%%%%%%%%%%%%%%%
Shown in Figure~1 is the temporal evolution of the mass accretion rate in unit of $M_\odot {\rm /sec}$ for the
models with $\beta = 0.5$ (blue), $\beta = 0.6$ (green) and $\beta = 0.75$ (red). The mass accretion rate is
measured at the innermost radius. The mass injection rate from the outer boundary is constant and common among
these models at $\dot{M}_{\rm out} \simeq 0.025M_\odot{\rm /sec}$. The parameters $\alpha_{\rm min} = 10^{-4}$, $\rm{Pm}_{\rm c} = 10$,
and $\mathcal{Q}_{\rm crit} = 5.0$ are also adopted commonly to these models.  

In the case with lower value of $\beta$ (see models with $\beta = 0.5$ and $0.6$), the disk gradually evolves to the
equilibrium state with the Pm-dependent $\alpha$ and then converges into the time-steady accretion stage  
compatible with the mass injection from the outer boundary. In contrast to them, the model with $\beta = 0.75$ shows
a time-dependent episodic mass accretion while the average value of $\dot{M}$ is almost same with other models.
In this model, as predicted from the linear theory, the inner region becomes secularly unstable because of the Pm-dependent $\alpha$ and as a result the gravitational instability is developed and makes the mass accretion sporadic.

%%%%%%%%%%%%%%%%%%%%%%%%%%%%%%%%%%%%%%%%%%%%%%%%%%%%%%%%%%%%%%%%%%%%
\begin{figure*}[tbp]
\begin{center}
\includegraphics[width=17cm,clip]{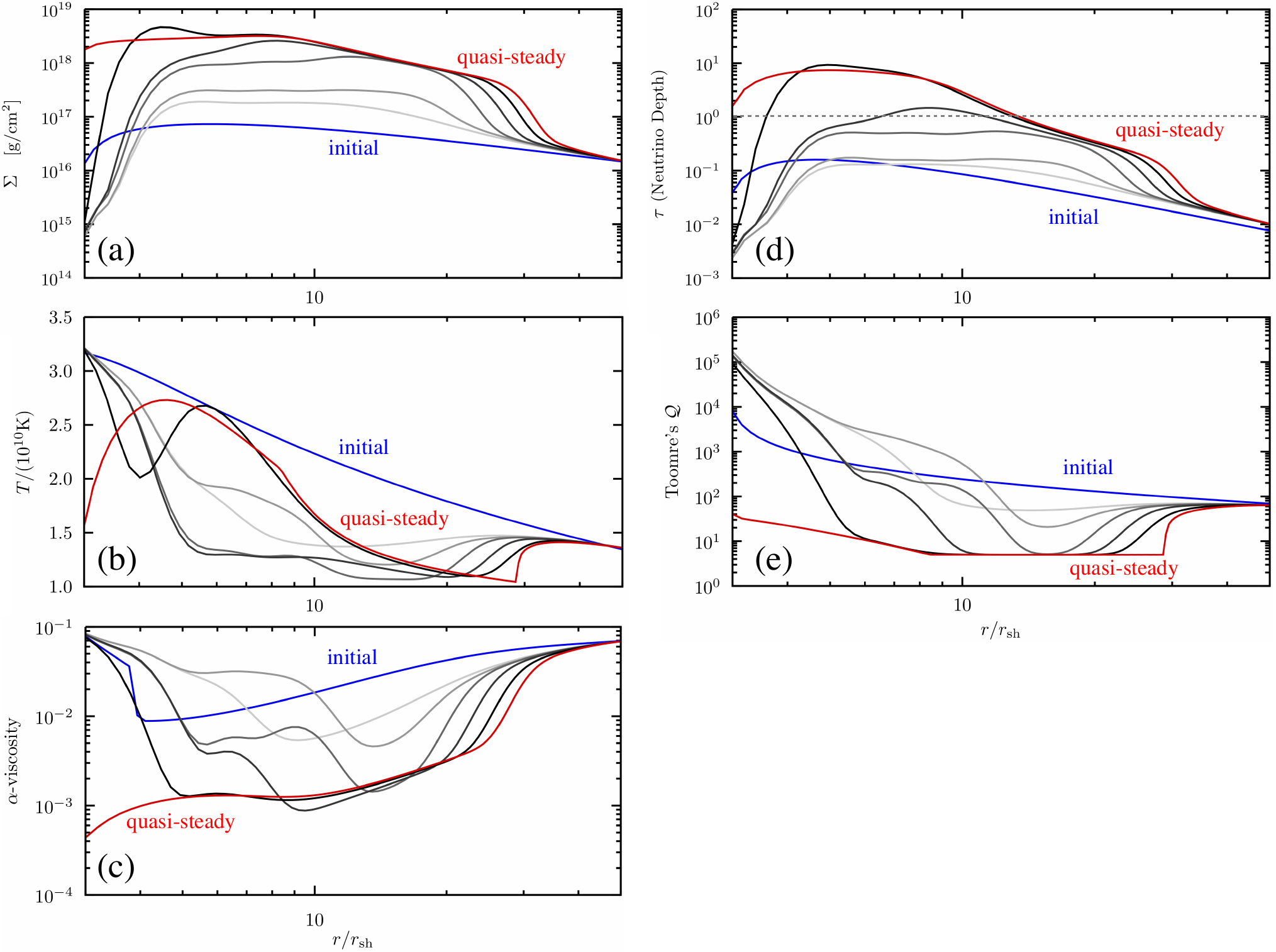}
\caption{Temporal evolution of radial profiles for (a) $\Sigma$, (b) $T$, (c) $\alpha$, (d) $\tau$, and (e) Toomre's $\mathcal{Q}$
  from the initial to the quasi-steady state.  The blue line corresponds to the initial equilibrium state consistent with constant $\alpha$,
  the red line is the quasi-steady in each panel, and the gray lines are in-between them, where the lines with darker tones depict the profiles at later times.  Here the parameters are set as $\beta=0.75$, $\alpha_{\rm min} = 10^{-4}$, $\rm{Pm}_{\rm c} = 10$,
and $\mathcal{Q}_{\rm crit} = 5.0$.} 
\label{fig2}
\end{center}
\end{figure*}
%%%%%%%%%%%%%%%%%%%%%%%%%%%%%%%%%%%%%%%%%%%%%%%%%%%%%%%%%%%%%%%%%%%%
Figure~2 shows the radial profiles of physical variables during the early evolution of the disk with $\beta = 0.75$ from the initial state 
to the quasi-steady state (around $t \simeq 4$ sec just before the initial spike), where the linear growth of the physical
variables are terminated and the mass accretion becomes almost steady when taking long-term average of it.  Panels~(a)--(f) are for
(a) $\Sigma$, (b) $T$, (c) $\alpha$, (d) $\tau$ (neutrino depth), and (e) Toomre's $\mathcal{Q}$.  In each panel, the blue line corresponds
to the initial equilibrium state with constant $\alpha$, the red line is the quasi-steady state and the gray lines are in-between
them, where the lines with darker tones depict the profiles at later times.

In the initial state the disk is geometrically thin and neutrino-thin, and the Toomre's $\mathcal{Q}$ is $\mathcal{O}(100)$ over the whole disk.  When applying the Pm-dependent $\alpha$-viscosity to it, $\alpha$ becomes lower in the inner part of the disk and then the secular
instability starts to grow there.  Since the secular instability leads to the mass accumulation, the surface density and thus the neutrino
depth also start to grow.  As a result, Toomre's $\mathcal{Q}$ decreases with time and finally reaches $\mathcal{Q}_{\rm crit}$ which causes
the gravitational instability, providing a drastic mass accretion. Once the gravitational instability grows, the inner part of the disk
repeats the mass accumulation stage and the gravitationally unstable stage, alternately. This would be a reason why the mass accretion
falls into the episodic mode in the secularly unstable disk with $\beta = 0.75$.

%%%%%%%%%%%%%%%%%%%%%%%%%%%%%%%%%%%%%%%%%%%%%%%%%%%%%%%%%%%%%%%%%%%%
\begin{figure*}[tbp]
\begin{center}
\includegraphics[width=17cm,clip]{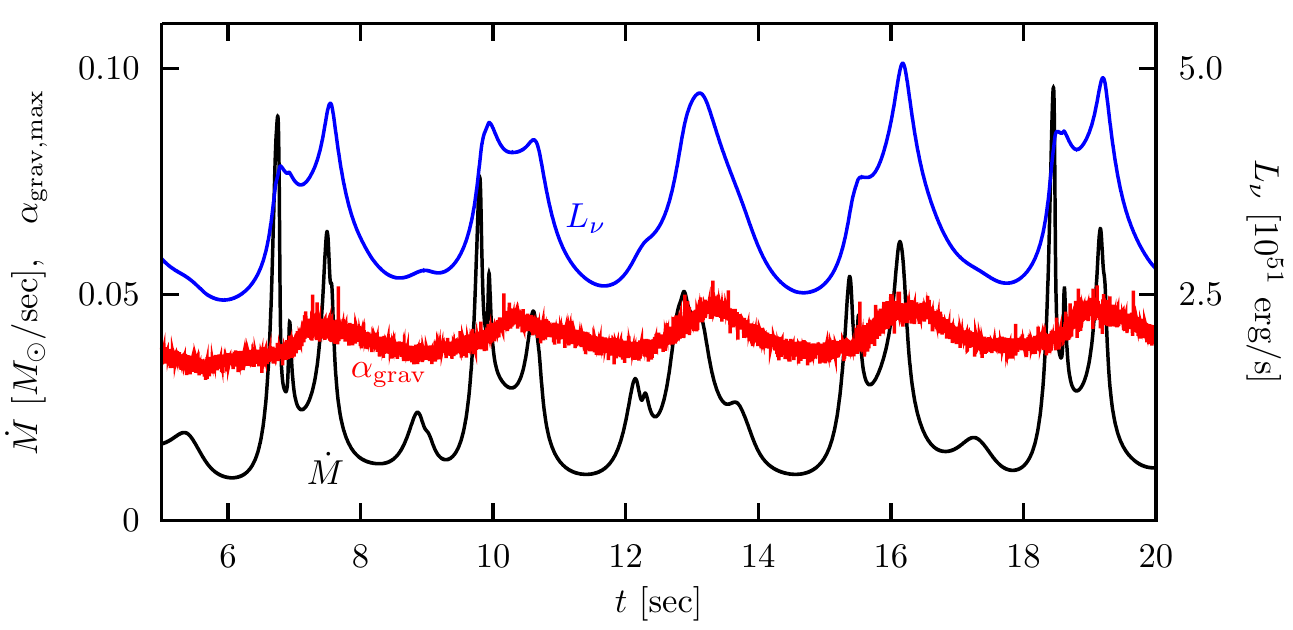}
\caption{Time-evolutions of $\dot{M}$ (black), the maximum value of $\alpha_{\rm grav}$ (red), and $L_\nu$ (blue)
  with focusing on the duration $6\ {\rm sec} \lesssim t \lesssim 20\ {\rm sec}$.  The parameters $\beta$, $\alpha_{\rm min}$, ${\rm Pm}_{\rm c}$, and $\mathcal{Q}_{\rm crit}$ are same as in Fig. 2.}
\label{fig3}
\end{center}
\end{figure*}
%%%%%%%%%%%%%%%%%%%%%%%%%%%%%%%%%%%%%%%%%%%%%%%%%%%%%%%%%%%%%%%%%%%%
Figure 3 depicts the time profiles of the $\dot{M}$ (black) and the maximum value of $\alpha_{\rm grav}$ (red) focusing on $6~{\rm sec}\lesssim t \lesssim 20~{\rm sec}$, showing that the mass accretion rate and the torque driven by gravitational instability are synchronized.  We can see that the cycle of mass accretion coincides well with the modulation cycle of the maximum value of $\alpha_{\rm grav}$ with relatively long period, $t_{\rm cycle} \sim 3~{\rm sec}$. 

This cycle period can be estimated from the redistribution time of the baryonic matter through the secularly unstable region with
the suppressed value of $\alpha$.  Since $\alpha$ is reduced to $10^{-4}$ in the inner secularly unstable region, the viscous
redistribution time $t_{\rm vis}$ of the matter through there is evaluated as
\begin{equation}
t_{\rm vis} \simeq \frac{r_{\rm SI}^2}{\nu_{\rm SI}} = \mathcal{O}(1)\ {\rm sec} \;,
\end{equation}
equivalent to the modulation period, where $r_{\rm SI} \sim \mathcal{O}(1)r_g$ is the radius of the secularly unstable region
and $\nu_{\rm SI}$ is the magnitude of the viscosity there, which is evaluated from the simulation result
as $\nu_{\rm SI} \simeq 10^{12}\ {\rm cm^2/sec}$. This implies that the global change of the disk structure is
regulated by the floor value of Pm-dependent $\alpha$, which is one of the control parameters in our simulation model. 

In addition to the longer period mass accretion cycle, we can find the shorter time-variability with a duration $\lesssim 1\ {\rm sec}$
in $\dot{M}$. Since the shorter wavelength mode of the secular instability has a faster growth rate, we can expect that 
such an additional component should be generated due to the development of smaller-scale structures by the localized secular instability.  Another possibility is that this component comes from the thermal instability occurring in the disk, whose timescale is generally much shorter than the viscous timescale (see Sec.4).

In Figure~3, the temporal evolution of the neutrino luminosity integrated over the whole disk, $L_\nu$, is also shown by a blue line.  As seen from this, $L_\nu$ also changes following the modulation of $\alpha_{\rm grav}$, suggesting again the change of the
global disk structure with this period. 

\subsection{Models with Power-law Decay of Mass Injection}
%%%%%%%%%%%%%%%%%%%%%%%%%%%%%%%%%%%%%%%%%%%%%%%%%%%%%%%%%%%%%%%%%%%%
\begin{figure*}[tbp]
\begin{center}
\includegraphics[width=17cm,clip]{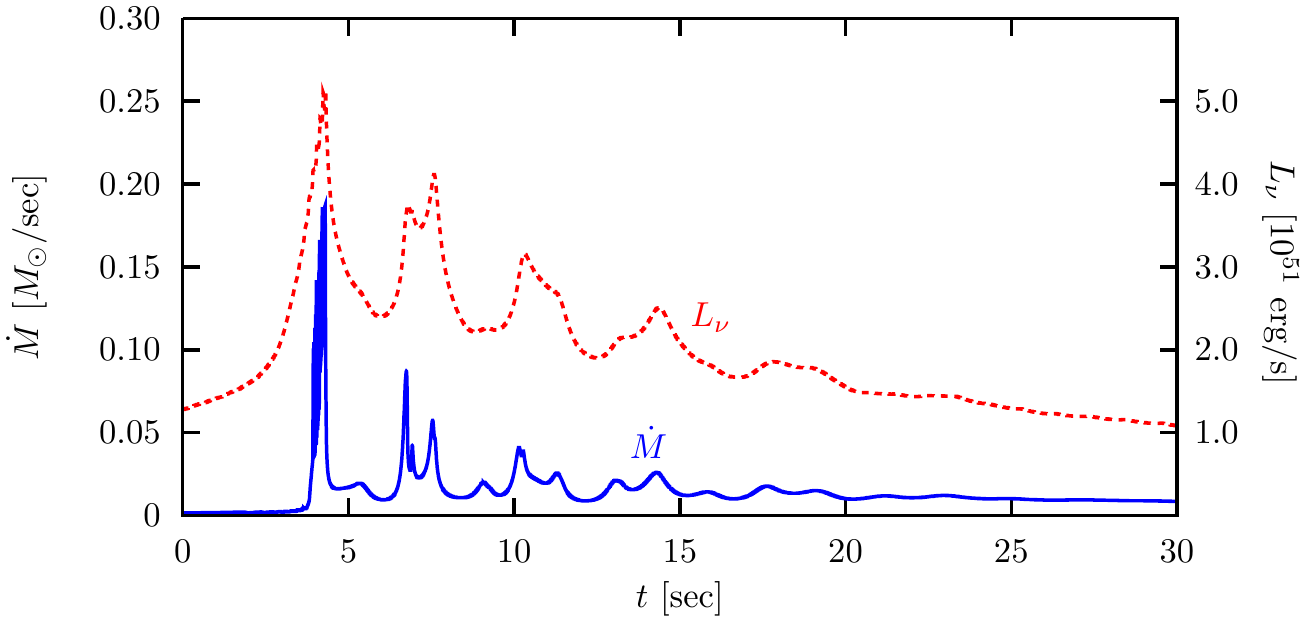}
\caption{Temporal evolution of $\dot{M}$ (red) and $L_\nu$ (blue) for the model with the same parameter sets as that in Figure~3
  but with decaying mass injection. } 
\label{fig4}
\end{center}
\end{figure*}
%%%%%%%%%%%%%%%%%%%%%%%%%%%%%%%%%%%%%%%%%%%%%%%%%%%%%%%%%%%%%%%%%%%%
The mass accretion rate from the outer boundary of the disk is expected to decrease with time in the realistic situation of the collapse of a massive star, which is prerequisite for the NDAF formation.  As the demonstration of such situations, under the condition of decaying mass injection,
we simulate the models with variety of parameter sets which give the secularly and gravitationally unstable disks. In these models, the
outer boundary condition is time-dependent and is controlled as
\begin{equation}
  \Sigma_{\rm out} =
  \left\{
\begin{array}{lr}
\Sigma_{{\rm out},0} & (t < t_{\rm ds})  \\
\Sigma_{{\rm out},0} (t/t_{\rm ds})^{-5/3} & (t_{\rm ds} \le t)
\end{array} \right.
\end{equation}
where $\Sigma_{{\rm out},0}$ is the initial surface density at the outer boundary, and $t_{\rm ds}$ is the start time of the decay.  We choose here $t_{\rm ds} = 5\ {\rm sec}$, which corresponds to the timing of the initial spike of the mass accretion.

Shown in Figure~4 is $\dot{M}$ (red) and $L_\nu$ (blue) for the model with the same parameter sets as that in Figure~3 but with decaying mass
injection.  Due to the decrease of $\Sigma_{\rm out}$, the duration of both $\dot{M}$ and $L_\nu$ becomes shorter and converges into $\sim 10$~sec.
Such a temporal behavior is reminiscent of the light curves of GRBs. Depending on the parameter sets, the temporal evolution of $\dot{M}$ exhibits 
various behaviors.

%%%%%%%%%%%%%%%%%%%%%%%%%%%%%%%%%%%%%%%%%%%%%%%%%%%%%%%%%%%%%%%%%%%%
\begin{figure*}[tbp]
\begin{center}
\includegraphics[width=17cm,clip]{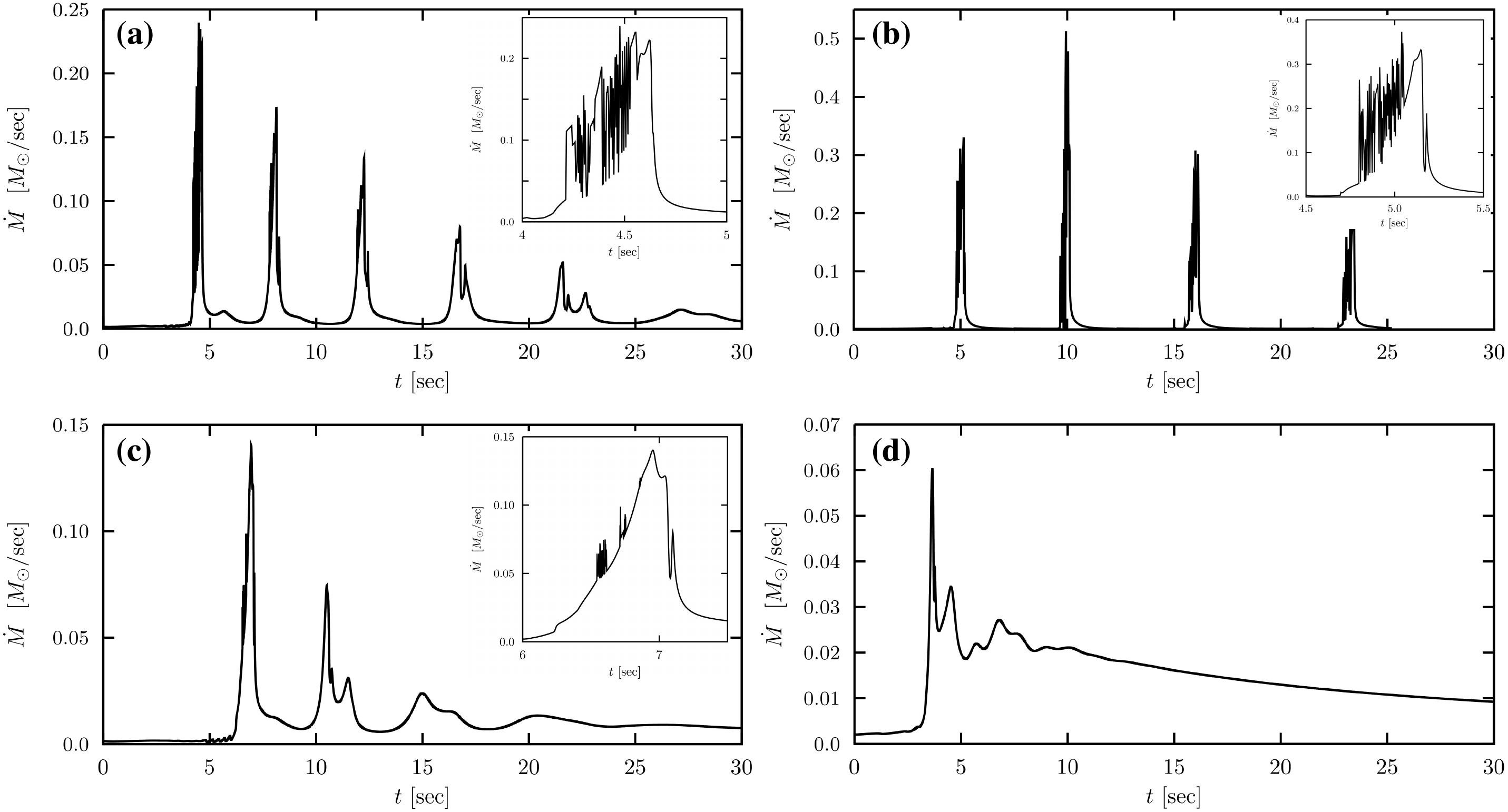}
\caption{Time-series of $\dot{M}$ for the models with different parameter sets: 
  (a) $\beta = 0.8,\ \alpha_{\rm min} = 10^{-4},\ \mathcal{Q}_{\rm crit} = 5.0,\ {\rm Pm}_{\rm c} = 10.0$,
(b) $\beta = 0.8,\ \alpha_{\rm min} = 10^{-5},\ \mathcal{Q}_{\rm crit} = 3.0,\ {\rm Pm}_{\rm c} = 10.0$, (c) $\beta = 0.9,\ \alpha_{\rm min} = 2\times 10^{-4},\ \mathcal{Q}_{\rm crit} = 3.0,\ {\rm Pm}_{\rm c} = 15.0 $, and (d) $\beta = 0.7,\ \alpha_{\rm min} = 2\times 10^{-4},\ \mathcal{Q}_{\rm crit} = 3.0,\ {\rm Pm}_{\rm c} = 10.0$
  The inset of each panels focus on the shorter-time variability in the first spike of $\dot{M}$. } 
\label{fig5}
\end{center}
\end{figure*}
%%%%%%%%%%%%%%%%%%%%%%%%%%%%%%%%%%%%%%%%%%%%%%%%%%%%%%%%%%%%%%%%%%%%
The temporal evolution of $\dot{M}$ for the models with different parameter sets are demonstrated in Figure~5.
Panels are for the models with (a) $\beta = 0.8,\ \alpha_{\rm min} = 10^{-4},\ \mathcal{Q}_{\rm crit} = 5.0,\ {\rm Pm}_{\rm c} = 10.0$,
(b) $\beta = 0.8,\ \alpha_{\rm min} = 10^{-5},\ \mathcal{Q}_{\rm crit} = 3.0,\ {\rm Pm}_{\rm c} = 10.0$, (c) $\beta = 0.9,\ \alpha_{\rm min} = 2\times 10^{-4},\ \mathcal{Q}_{\rm crit} = 3.0,\ {\rm Pm}_{\rm c} = 15.0 $, and (d) $\beta = 0.7,\ \alpha_{\rm min} = 2\times 10^{-4},\ \mathcal{Q}_{\rm crit} = 3.0,\ {\rm Pm}_{\rm c} = 10.0$. The inset of each panels focus on the shorter-time variability in the first spike of $\dot{M}$.

We can see that with the smaller value of $\alpha_{\rm min}$, the modulation cycle becomes longer (see, panels~(a)--(c)).  While the time-decaying behavior is common among them, in the case of model~(d) with $\beta = 0.7$ which is located near the critical value, the cyclic behavior almost disappears, showing one longer
humped structure.  Of course, we understand that the time-scale of the variability should be studied by more realistic MHD models of the NDAF because the
various important effects lacks in our 1D simulation, such as advection effects, pressure gradient effect, and non-linear interaction between them.
Even though our model is far from the realistic one, such a variety of temporal behavior of the mass accretion rate is reminiscent of the diversity of
the light curve of GRBs. 

%%%%%%%%%%%%%%%%%%%%%%%%%%%%%%%%%%%%%%%%%%%%%%%%%%%%%%%%%%%%%%%%%%%%
\section{Discussion}
\subsection{Disk Evolution on $\dot{M}$--$\Sigma$ plane}
The episodic mass accretion process seen in our simulations naturally provokes a question 
``how an accretion disk changes its state on $\dot{M}-\Sigma$ or $T-\Sigma$ plane in this model ?".
%how our disk model changes the state on $\dot{M}$--$\Sigma$ or $T$--$\Sigma$ plane ?".  
To answer it, we discuss %the path of the disk evolution 
the evolution path of a disk on these planes here. 
In the following, we cast a spotlight on the model demonstrated in Figure 5(b) as a typical one.

Figure 6(a) shows the time variation of $\dot{M}$ %with 
focusing on the period %time-span 
around the secondary burst ($t \sim 10$ sec) for that model. 
The thick green line, that denotes the low resolution time-series, 
is superimposed on the gray curve that denotes highly time-resolved data. 
Shown in the panels (b)~and~(c) in Figure 6 are the disk evolution on $\dot{M}$--$\Sigma$ and $T$--$\Sigma$ planes at the disk 
inner edge, $r = 3r_g$. The reference points $1$--$12$ in the panels (b) and (c) correspond to the states at the reference 
times $1$--$12$ in the panel (a), respectively.

%In the earlier thinner disk stage, 
In the earlier stage when a disk is optically thin with respect to neutrinos,
both the $\dot{M}$ and $T$ are lower because of the low $\alpha$-value 
suppressed at the inner region due to its $Pm$-dependence (states~$1$--$2$). Since the $\dot{M}$ is higher in the outer region, 
the baryonic matter is accumulated in the inner part during this stage and thus the $\Sigma$ gradually increases with time 
(states~$3\rightarrow 5$). 
%The higher the $\Sigma$, the $T$ also becomes higher because of the increase of the viscous heating, 
When $\Sigma$ gets larger, $T$ becomes higher because viscous heating becomes more effective, 
resulting in the increase of $\alpha$ and thus $\dot{M}$. The growths of the $\Sigma$ and $T$ are terminated when the 
Toomre's $\mathcal{Q}$ reaches the critical value at the inner region. Then, the disk enters the bursting mass accretion stage with 
high turbulent viscosity ($\because \alpha_{\rm grav} > \alpha$) (state~$6$), and finally returns to the thinner and cooler state 
(states~$6\rightarrow 12$) similar to the starting point.

Although one typical mode is focused here, the other models also show the similar behavior qualitatively on 
the $\dot{M}$--$\Sigma$ plane. To summarize, our results thus show that the evolution of an NDAF in our model has the limit-cycle 
property which is reminiscent of that in the other astrophysical disk models, such as dwarf novae (e.g. \citealp{1991PASJ...43..147H}). 

%\textcolor{red}{The evolution of our NDAF model on $\dot{M}$--$\Sigma$ is reminiscent of the limit cycle behavior of the disk  
%for the case of the dwarf nova (e.g. \citealp{1991PASJ...43..147H}). The though the dominant mechanism may be different between them. } 
%%%%%%%%%%%%%%%%%%%%%%%%%%%%%%%%%%%%%%%%%%%%%%%%%%%%%%%%%%%%%%%%%%%%
\begin{figure}[tbp]
\begin{center}
\includegraphics[width=8cm,clip]{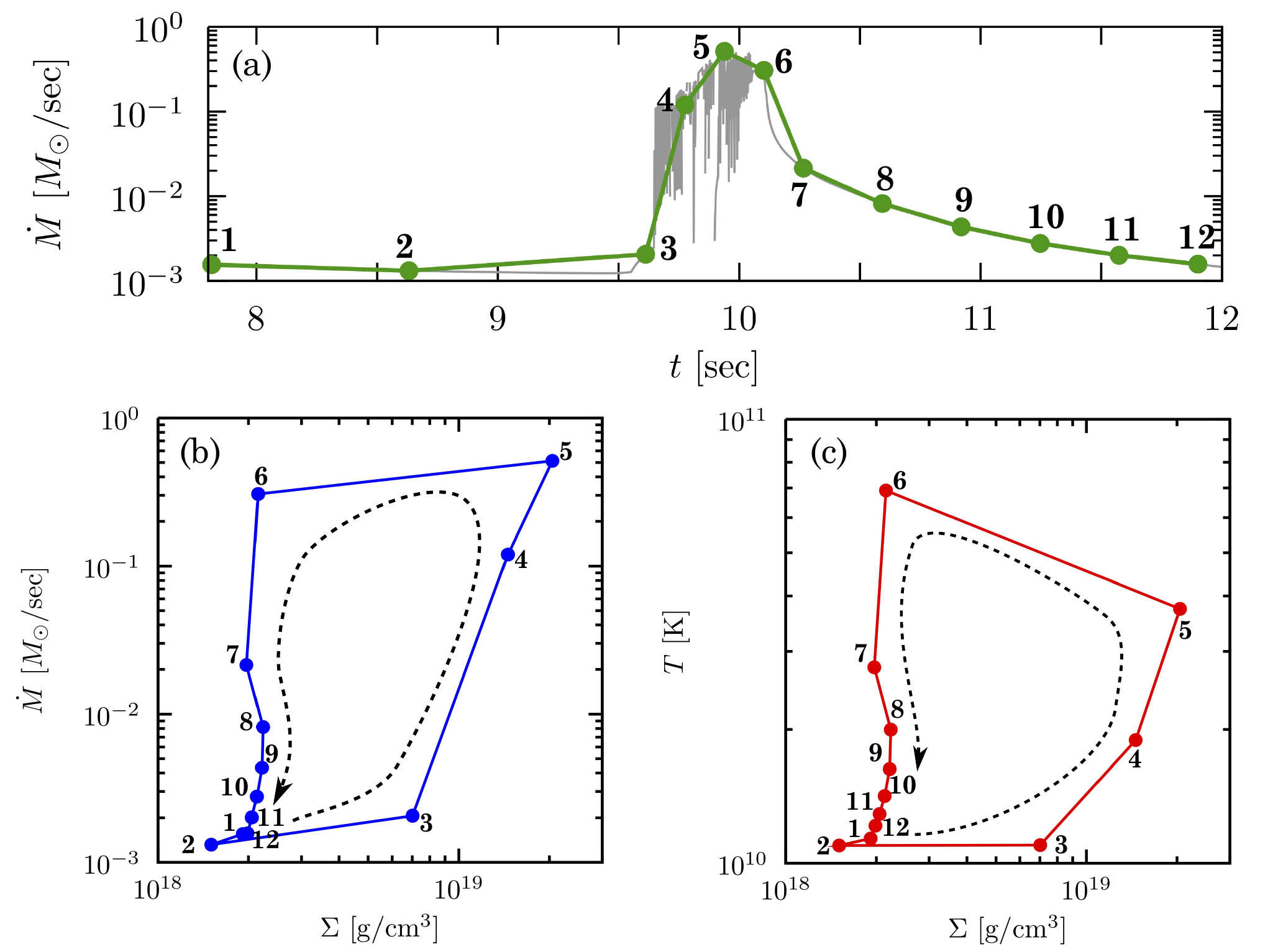}
\caption{Path of the disk evolution. (a)the time variation of $\dot{M}$ around the secondary burst ($t\sim 10$sec) 
for the model in Figure 5(b). The thick green line, which denotes the low resolution time-series, is superimposed on the gray curve 
that denotes highly time-resolved data. Panels (b)~and~(c) are the disk evolution on $\dot{M}$--$\Sigma$ and $T$--$\Sigma$ planes 
at $r = 3r_g$ (inner edge). The reference points $1$--$12$ in the panels (b) and (c) correspond to the states at the reference 
times $1$--$12$ in the panel (a), respectively.}
\label{fig6}
\end{center}
\end{figure}
%%%%%%%%%%%%%%%%%%%%%%%%%%%%%%%%%%%%%%%%%%%%%%%%%%%%%%%%%%%%%%%%%%%%
\subsection{Application to Astrophysical Objects}
The origin of short-term variability in the prompt emission of GRBs is still unclear.  One of the most promising mechanisms is the accretion disk instability, which has been discussed to interpret the flaring behavior of X-ray binaries and dwarf novae, etc.  Several models of the disk instability in a hyperaccretion flow have been proposed \citep{2007ApJ...664.1011J,masada+07,2012MNRAS.419..713K,2013ApJ...777L..15K,2015PASJ...67..101K}.  However, most of these models rely on the prescription with a constant $\alpha$-viscosity, and since the mass accretion process is governed by the angular momentum transportation driven by the growth of MRI inside the disk, it is necessary to follow the fundamental MHD process in the neutrino-emitting gas in order to investigate the variability occurring in the hyperaccretion flow.  Only \cite{masada+07} discussed the possibility that the angular momentum transport is suppressed because MRI in a neutrino-thick accretion flow would be suppressed, which would cause the gravitational instability in the inner disk.  Such a mechanism works only if the accretion rate is so high that the accretion flow becomes dense and the optical depth with respect to neutrinos is larger than unity.  We have applied the prescription in which the $\alpha$-viscosity is not constant but depends on the magnetic Prandtl number, ${\rm Pm}=\nu/\eta$, to a hyperaccretion flow for the first time.  The $\alpha$ parameter represents the saturation level of MRI in the disk, which would be impacted by magnetic reconnection rate of the MRI-amplified magnetic field.  The higher viscosity suppresses efficiently the small scale turbulent structure, sustaining the larger size of the turbulent vortices and thus the larger size of the frozen-in magnetic structure.  Since magnetic reconnection rate is expected to be smaller with the larger size of the turbulent magnetic structure and the smaller magnetic resistivity, the saturation level of the MRI-driven turbulence would become higher with the increase of the Pm.  We make use of the results of \cite{2008MNRAS.391..922R} to express the magnetic Prandtl number in a dense gas, where electrons are relativistically degenerate, as an analytic function of temperature and density.  By deriving the thermal equilibrium solution of a hyperaccretion flow which is cooled efficiently via neutrino emission, assuming $\alpha$ depends on the magnetic Prandtl number as $\alpha\propto {\rm Pm}^{\beta}$, we show that when the exponent $\beta$ takes a value in a certain range a neutrino-cooled accretion flow becomes secularly unstable.  To investigate what will occur in a hyperaccretion flow when this condition is fulfilled we perform one-dimensional simulations of a hyperaccretion flow with Pm-dependent $\alpha$-viscosity and show that if the index $\beta$ is within this range the mass would be accumulated in the inner part of the accretion flow because the angular momentum transportation due to the turbulence driven by MRI would be suppressed as the surface density grows.  This mass accumulation makes an accretion flow gravitationally unstable at some stage and intermittent mass accretion onto a black hole due to the gravitational torque would occur.  This may cause the ejection of inhomogeneous jets, which can be the origin of the short-term variability of GRBs. 

As can be seen in Fig. 1 and 3, the variability of mass accretion can be decomposed into long-term cycles whose timescale is $\sim \mathcal{O}(1)~{\rm sec}$ and short-term oscillations whose timescale is $\lesssim 0.1-1~{\rm sec}$.  Actually it has been suggested that a GRB lightcurve contains two components whose timescales are different from each other (e.g. \citealp{2006A&A...447..499V}), which may be produced by the mechanism shown in our simulations.  The long-term oscillation corresponds to the viscous redistribution of disk material which is regulated by the floor value of Pm-dependent $\alpha$, while it is not obvious what makes the latter oscillation.  One of the possibilities is that it comes from the shorter wavelength modes of the secular instability, which have a faster growing rate.  Another possibility is that the thermal instability, whose timescale is generally much shorter than the viscous timescale.  Let us discuss the conditions for thermal instability to occur in a hyperaccretion flow.  We can check if the disk is thermally stable or not by comparing the temperature dependence of $Q^+$ and $Q^{-}$ in the thermal equilibrium, keeping the surface density $\Sigma$ constant \citep{1976MNRAS.177...65P}.  In the optically-thin regime, from Eqs.(\ref{hydrostatic}), (\ref{energybalance}), (\ref{alpha}), and (\ref{prandtlrhotemp}), we have the temperature dependence of the heating rate as $Q^+ \propto T^{5\beta/2+1}$, and from Eq. (\ref{urcathin}) we have $Q^- \propto T^6$.  Therefore, an optically-thin disk becomes thermally unstable when $\beta >2$, and we can see that this condition contradicts with Eq.(\ref{thinunstable}).  In other words, an optically-thin disk cannot become secularly unstable and thermally unstable at the same time.  In the optically-thick disk, however, the situation changes completely.  Using Eq.(\ref{urcathick}), we have $Q^- \propto T^2$ and the condition that an optically-thick disk becomes thermally unstable can be described as $\beta > 2/5$.  This is identical with the condition that an optically-thick disk becomes secularly unstable, Eq. (\ref{thickunstable}).  Therefore, in the optically-thick regime, a disk that is secularly unstable is always thermally unstable.  As can be seen in Fig.2 the innermost region of the disk becomes optically-thick as the secular instability develops, and so it should be thermally unstable at the same time.  The thermal timescale in an accretion disk is described as $t_{\rm th}\sim 1/(\alpha \Omega)$, while the viscous timescale is described as $t_{\rm vis}\sim 1/(\alpha \Omega)(r/H)^2$.  Since the disk is geometrically thin (i.e. $H \ll r$), the thermal timescale is much shorter than the viscous timescale.  This means that when a disk is in the optically-thick regime, the thermal instability develops faster than the secular instability.  As shown in Fig.3 the time profiles of mass accretion rate have the shorter-time variability with duration $\lesssim 1~{\rm sec}$.  Since the disk simulated here is in the optically-thick regime at the evolved state, it is not only secularly unstable but also thermally unstable.  Thus, the short-time variability may originate essentially from the thermal instability whose timescale is much shorter than the viscous timescale (e.g., \citealp{2007ApJ...664.1011J}).  It is beyond the scope of this paper to specify the origin of this short-term variability but is one of the main targets of our future work with ``more realistic'' multi-dimensional time-dependent model.

The viscosity in a hyperaccretion flow is important not only in the context of the central engine of GRBs but also in the context of the electromagnetic counterpart of gravitational waves from binary neutron star mergers or black hole - neutron star mergers.  From the remnant accretion disks of compact binary mergers, one can expect the disk outflow driven by viscous heating \citep{2008MNRAS.390..781M, 2009MNRAS.396..304M, 2013MNRAS.435..502F, 2014MNRAS.441.3444M, 2015ApJ...813....2M, 2017PhRvL.119w1102S, 2017MNRAS.472..904L, 2018ApJ...858...52S, 2017ApJ...846..114F, 2018ApJ...860...64F, 2018arXiv180800461F, 2018arXiv180911163R}.  For example, \cite{2018ApJ...860...64F} investigated the mass ejection from a torus surrounding a massive neutron star, which is formed after the merger of binary neutron stars by general relativistic neutrino-radiation hydrodynamics simulations.  They discussed the importance of mass ejection due to the viscous heating in the torus, using $\alpha$-prescription to model the shear viscosity.  However, they assume that $\alpha$ is a constant ($\sim 0.01-0.04$) in a whole torus.  As we have shown above, due to the degeneracy of electrons the $\alpha$-parameter may no longer be constant in the innermost region of a hyperaccretion flow.  In fact, we can see in Fig.2 (c) that the $\alpha$-parameter largely depends on radius.  Therefore, the behavior of mass ejection due to the viscosity can change if taking into account this effect, which is beyond the scope of this paper.

\section{Summary}
We investigate the stability of a hyperaccretion flow, which is considered to be the central engine of GRBs.  Especially we take into account the variation of the viscosity parameter $\alpha$ depending on the magnetic Prandtl number, which has recently been implied by several MHD simulations, in the context of a hyperaccretion flow for the first time.  We have shown that a hyperaccretion flow becomes secularly unstable when $\alpha$ is proportional to the magnetic Prandtl number with an exponent $\beta$ within a certain range, which is consistent with the value inferred from recent MHD simulations.  Once a hyperaccretion flow becomes secularly unstable, the surface density of the inner region increases due to the mass accumulation.  As a result, the inner region of the accretion flow becomes gravitationally unstable, and intense mass accretion would occur induced by non-axisymmetric pattern in the accretion flow.  Since such an accretion process can occur repeatedly as long as matter is sufficiently supplied from the outer region, we can expect sporadic mass accretion in this model, which may account for the short-term variability observed in the prompt emission of GRBs.  We also perform one-dimensional simulations of a hyperaccretion flow assuming the Pm-dependence of the $\alpha$-parameter, and we can reproduce the intermittent behavior of an accretion flow which is secularly unstable.
\ \\

This work was supported by JSPS KAKENHI Grants No. 18K03700, No. 18H04444, and No. 18H01212.  NK acknowledges the support from the Hakubi project at Kyoto University.  Numerical computations were in part carried out on Cray XC50 at Center for Computational Astrophysics,
National Astronomical Observatory of Japan.

\end{document}